\documentclass[fleqn,10pt]{wlscirep}
\usepackage[utf8]{inputenc}
\usepackage[T1]{fontenc}
\usepackage{lineno}
\usepackage{amsmath}
\usepackage{amssymb}
\usepackage{newfloat}

\usepackage[marginal]{footmisc}
\usepackage{multirow}
\usepackage{threeparttable}
\DeclareFloatingEnvironment[name={Supplementary Figure}]{suppfigure}
\DeclareFloatingEnvironment[name={Supplementary Table}]{supptable}

\begin{document}

\renewcommand{\thefootnote}{}
\newcommand{\xmli}[1]{{\color{red}{[XM: #1]}}}
\newcommand{\yq}[1]{{\color{red}{#1}}}
\newcommand{\cgs}[1]{#1}
\newcommand{\cgss}[1]{#1}
\newcommand{\etal}{\textit{et al.}}
\newcommand{\nickname}{SSR-KD}
\newcommand{\loss}{\mathcal{L}}

\title{\cgss{Real-Time Reconstruction of 3D Bone Models via Very-Low-Dose Protocols}}

\author[1,$\dag$]{Yiqun Lin}
\author[3,$\dag$]{Haoran Sun}
\author[2]{Yongqing Li}
\author[3]{Rabia Aslam}
\author[4]{Lung Fung Tse}
\author[5]{Tiange Cheng}
\author[6]{Chun Sing Chui}
\author[3]{Wing Fung Yau}
\author[3]{Victorine R. Le Meur}
\author[3]{Meruyert Amangeldy}
\author[5]{Kiho Cho}
\author[7]{Yinyu Ye}
\author[8]{James Zou}
\author[2,*]{Wei Zhao}
\author[1,*]{Xiaomeng Li}
\affil[1]{Department of Electronic and Computer Engineering, The Hong Kong University of Science and Technology, Hong Kong SAR}
\affil[2]{Department of Physics, Beihang University, Beijing, China}
\affil[3]{Koln 3D Technology (Medical) Limited, Hong Kong SAR}
\affil[4]{Union Hospital, Hong Kong SAR}
\affil[5]{Dental Materials Science, Division of Applied Oral Sciences and Community Dental Care, Faculty of Dentistry, The University of Hong Kong, Hong Kong SAR}
\affil[6]{Department of Orthopaedics and Traumatology, The Chinese University of Hong Kong, Hong Kong SAR}
\affil[7]{Department of Management Science and Engineering, Stanford University, Stanford, CA, USA}
\affil[8]{Department of Biomedical Data Science, Stanford University, Stanford, CA, USA}
\affil[*]{Correspondence: eexmli@ust.hk, zhaow20@buaa.edu.cn}
\affil[$\dag$]{These authors contributed equally to this work}

\begin{abstract}

Patient-specific bone models are essential for designing surgical guides and preoperative planning, as they enable the visualization of intricate anatomical structures. However, traditional CT-based approaches for creating bone models are limited to preoperative use due to the low flexibility and high radiation exposure of CT and time-consuming manual delineation. Here, we introduce \cgs{Semi-Supervised Reconstruction with Knowledge Distillation (SSR-KD)}, a fast and accurate AI framework to reconstruct high-quality bone models from biplanar X-rays in 30 seconds, with an average error under 1.0 mm, eliminating the dependence on CT and manual work. Additionally, high tibial osteotomy simulation was performed by experts on reconstructed bone models, demonstrating that bone models reconstructed from biplanar X-rays have comparable clinical applicability to those annotated from CT. Overall, our approach accelerates the process, reduces radiation exposure, enables intraoperative guidance, and significantly improves the practicality of bone models, offering transformative applications in orthopedics.


\end{abstract}

\flushbottom
\maketitle

\newpage
\section*{Introduction}

\begin{figure}[t]
\centering
\includegraphics[width=1.0\linewidth]{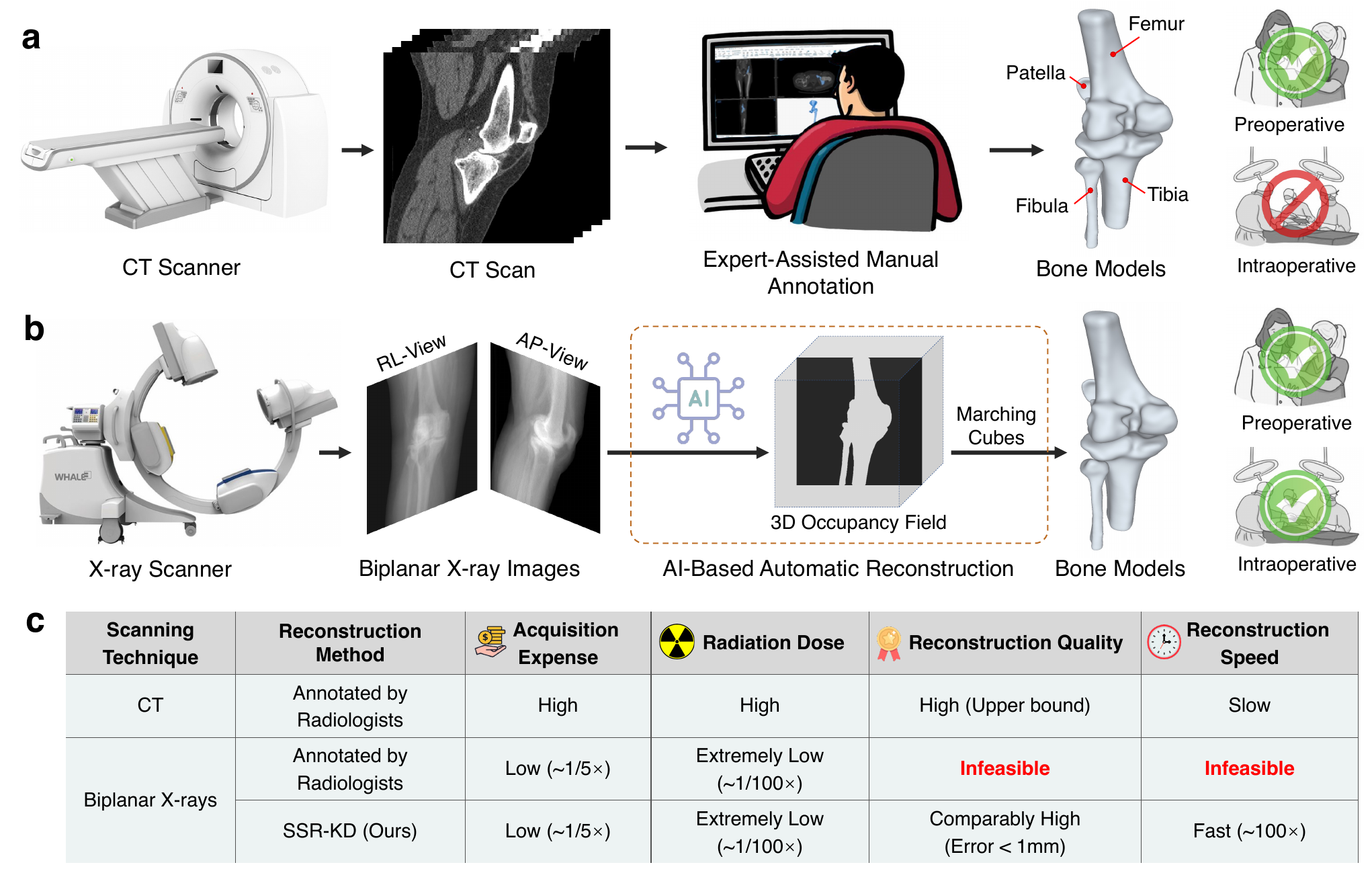}
\caption{
\textbf{Comparison of conventional reconstruction and our proposed solution.}
\textbf{a:} Conventional CT-based reconstruction. Traditionally, the CT scan of a patient should be collected first, and then the expert will operate commercial software like Mimics to delineate 3D bone models from the CT scan.
\textbf{b:} Our proposed automatic reconstruction from biplanar X-rays. The approach requires only two X-rays of the patient; then, the deep reconstruction network SSR-KD will automatically estimate the occupancy field; finally, Marching Cubes~\cite{lorensen1987marching} is applied to extract underlying iso-surfaces and reconstruct 3D bone models.
\textbf{c:} Comprehensive comparison of different reconstruction pipelines. Our SSR-KD uses low-cost, low-radiation biplanar X-rays to automatically create 3D bone models. The reconstruction is fast enough to be used during surgery and produces results comparable in quality to CT scans without needing expert assistance.}
\label{fig:pipeline}
\end{figure}

\noindent
Orthopedic procedures are performed at a significant scale each year. In 2022 alone, there were approximately 18.5 million orthopedic procedures conducted in the United States, with a global estimate reaching around 28.3 million~\cite{GOSMR,globaldata}. Within these procedures, patient-specific 3D bone models play a critical role in both preoperative and intraoperative stages. These models provide a comprehensive visualization of the intricate anatomical structure, enabling surgeons to develop a precise understanding, perform accurate measurements, plan procedures, and simulate various scenarios prior to surgery~\cite{li2019surgical,woo2020three}.
During the intraoperative stage, the utilization of 3D bone models enables surgeons to acquire clear and detailed observations of the 3D structures. This enhanced visualization significantly enhances surgical accuracy, ultimately leading to improved surgical outcomes~\cite{portnoy2023three, kumar2021does}.
In the preoperative stage, the typical method for obtaining patient-specific bone models involves undergoing a computed tomography (CT) scan. Subsequently, clinicians manually delineate the bone models from the CT volumes using commercial software, as illustrated in Fig.~\ref{fig:pipeline}a.

Obtaining high-quality bone models relies on 3D imaging devices such as CT scanners (e.g., O-Arm, Spiral CT), while conventional CT-based reconstruction (Fig.~\ref{fig:pipeline}a) to generate bone models suffers from notable limitations:
\textbf{(1) High radiation exposure:}
\cgs{CT scans typically require hundreds of X-ray projections to create a detailed 3D image~\cite{villarraga2020effect}, which exposes the patient to a significantly high radiation dose. This level of radiation exposure is a notable concern in clinical practice~\cite{brenner2007computed, miglioretti2013use}, particularly for vulnerable populations such as pediatric patients and pregnant women~\cite{pearce2012radiation, lee2004diagnostic}.}
\textbf{(2) Low flexibility:} 
CT scanners like Spiral CT are often bulky and occupy a large amount of space, limiting their use in intraoperative scenarios. \cgs{Although intraoperative 3D imaging systems like the O-Arm can be used during surgery, their integration involves a notable time investment, taking approximately 10–15 minutes for sterile draping and machine positioning~\cite{patil2012pedicle}, with an additional 13 minutes often required for navigated procedures~\cite{sari2024reduced}. Furthermore, radiation safety protocols require non-essential staff to exit the room during the 3D scan, disrupting the surgical workflow and limiting the frequency of 3D intraoperative imaging.}
\textbf{(3) Time-consuming reconstruction:} Even with the assistance of commercial software for generating bone models from CT scans, the process requires several hours of manual work by experts to process the CT volumes and accurately delineate the bone models. 

Considering these limitations, it is not feasible to perform conventional CT-based reconstructions in real-time during surgery. The objective of this study is to reconstruct high-quality 3D bone models directly from fluoroscopy images (specifically, biplanar X-rays; refer to Fig.~\ref{fig:pipeline}b). On one hand, this allows for the acquisition of 3D bone models while significantly minimizing radiation exposure. On the other hand, the use of fluoroscopy equipment \cgs{(e.g., C-Arm, G-Arm, and O-Arm systems)} offers enhanced flexibility, enabling utilization both in preoperative and intraoperative scenarios, and proves to be cost-effective.

To reconstruct bone models from X-rays, statistical models were proposed in early years to estimate the statistical shape model~\cite{sarkalkan2014statistical, baka2012statistical, zheng20092d, thusini2020uncertainty, lamecker2006atlas} (SSM) or statistical shape intensity model~\cite{klima2016intensity, ehlke2013fast, zheng2011personalized, sadowsky2007deformable, yao2003assessing} (SSIM) from a set of labeled examples, which can be regarded as an average shape of the bone model. These models can be further optimized and deformed to match the contour of bones in X-rays. Recently, deep learning techniques have shown remarkable applications in medical image translation~\cite{yu2024spatial, bellemo2024optical, chen2024translating, vcavojska2020estimating, jiang2022deep, preetha2021deep, magdy2024bone, wu2021machine}. Several studies have demonstrated the feasibility of using deep learning for bone model reconstruction from X-rays, utilizing approaches based on \cgs{convolutional neural networks~\cite{shiode20212d, kasten2020end} (CNNs)}, or incorporating CNNs with statistical models~\cite{aubert2019toward, kim20193d}. However, the clinical practicality of prior X-ray-based reconstruction methods is unknown or even poor due to several factors: 
\textbf{(1) Low quality:} Existing methods are limited to generating coarse and low-quality bone models from X-rays because SSM/SSIM-based methods are difficult to optimize, while learning-based methods require high computational costs, resulting in low reconstruction resolution.
\textbf{(2) Heavy dependency on labeled data:} Existing methods are data-driven, where the reconstruction quality is highly dependent on the number of labeled bone models used to estimate average bone models for patient-specific deformation or train the deep networks. 
\textbf{(3) Lack of evaluation to validate clinical applicability:} Existing evaluations are limited to quantitative metrics and qualitative comparisons, which hardly provide insights into the clinical significance and effectiveness of the reconstructed bone models. 

In this study, we collected paired training data consisting of biplanar X-rays (inputs) and 3D bone models (as the ground truth for supervision) manually annotated from corresponding CT volumes. Then, we proposed a new reconstruction framework 
\underline{\textbf{S}}emi-\underline{\textbf{S}}upervised \underline{\textbf{R}}econstruction with \underline{\textbf{K}}nowledge \underline{\textbf{D}}istillation (\nickname{})
to perform fast and accurate bone model reconstruction from biplanar X-rays. \cgs{A comprehensive comparison of the CT-based solution and our proposed SSR-KD is illustrated in Fig.~\ref{fig:pipeline}c.}
Instead of presenting bone models as mesh or volumes, we regard the bone model as an implicit field~\cite{park2019deepsdf, mescheder2019occupancy} (specifically, occupancy field) to improve learning efficiency and achieve better reconstruction quality and higher resolution.
Considering the time-consuming labeling, we additionally utilized unlabeled data for training and leveraged semi-supervised learning and knowledge distillation to further improve the reconstruction quality.
We quantitatively and qualitatively evaluated our method on a clinical dataset, showing superior reconstruction quality (resolution $\geq$256$^\text{3}$ and average error $\leq$1.0 mm) and fast reconstruction speed ($<$30 seconds).
More importantly, we further conducted clinical applicability evaluation by performing \cgs{high tibial osteotomy (HTO)} simulation, wherein experienced surgeons and medical professionals evaluated the fitting, stability, and accuracy of surgical guides designed based on 2-view reconstructed and the corresponding CT-based annotated bone models. The results confirmed that surgical guides designed based on bone models reconstructed from biplanar X-rays using our \nickname{} exhibit comparable accuracy to those designed using bone models manually annotated from CT, which validated the feasibility of our proposed algorithm in generating bone models suitable for clinical practice.

\section*{Results}

\subsection*{Overall Framework}
%
We formulated the reconstruction as the estimation of the 3D occupancy field, where 3D points inside the bone surface have an occupancy value of $+1$, and points outside the bone surface have an occupancy value of $0$. 
%
%
As shown in Fig.~\ref{fig:pipeline}b, given anterior-posterior (AP) and right-left (RL) X-ray images as the input, our proposed approach predicts the occupancy values of points in the 3D space to form the occupancy field; then, the surfaces are extracted from the occupancy field using Marching Cubes~\cite{lorensen1987marching}.
%
%
The deep neural network was designed to learn a mapping function from a 3D point to a scalar occupancy value; see \textit{Reconstruction Networks} in \textit{Methods} section. Specifically, the encoder network first extracts the semantic features from input images; then, for a 3D point, the pixel-aligned features are queried from 2D feature maps by its spatial coordinates using bilinear interpolation. Finally, fully connected layers are used to predict the occupancy value. \cgss{The proposed approach achieved a reconstruction time of approximately 25.2 seconds (network inference: 20.8s, Marching Cubes: 1.5s, post-processing: 2.9s),} significantly faster than the time required for manually annotating bone models from CT.

%
In this work, given biplanar (2-view) X-ray images, we aimed to reconstruct four bone models of a single leg, including patella, femur, fibula, and tibia. Accurate bone models were manually extracted from the corresponding CT scan as the ground-truth supervision during the training process. According to our common sense, a large amount of annotated data is required to train a deep neural network. On one hand, instead of collecting paired and registered X-ray and CT from the same patient, we follow previous works~\cite{shen2019patient, shiode20212d, reyneke2018review} to digitally generate X-ray projections from CT scans. On the other hand, manual annotation of a single leg usually requires approximately 4 hours for an experienced orthopedic expert, which is time-consuming and expertise-demanded. \cgs{Hence, we utilized 70 ($\sim$12.6\%) labeled and 485 ($\sim$87.4\%) unlabeled data for semi-supervised training.} Details about the dataset collection and network design \& training are provided in the Methods section.

\subsection*{Quantitative and Qualitative Evaluation}

For quantitative evaluation, we followed previous works on bone reconstruction~\cite{klima2016intensity, chenes2021revisiting} to use average symmetric surface distance (ASSD, mm) and Hausdorff distance (HD, mm) to measure the reconstruction error between predicted bone models and ground-truth models (annotated from CT). We also reported the Dice similarity coefficient (DSC, \%) based on voxelized 3D bone models. The comprehensive formulation of the aforementioned metrics is provided in \textit{Methods} section; refer to the \textit{Metrics for Quantitative Evaluation}. An independent testing dataset consisting of 40 data pairs was used for evaluation. Quantitative results in Tab.~\ref{tab:results} demonstrate that our proposed SSR-KD achieved superior performance than previous studies with an average ASSD below 1~mm. In particular, the surface error (ASSD) of femur and tibia is lower than 0.8~mm, which is very close to the average voxel-wise spacing (0.68~mm) of CT scans. We observed that the shapes of patella and fibula are partially or completely occluded in input views, resulting in higher reconstruction error than femur and tibia.

\begin{figure}
\centering
\includegraphics[width=1.0\linewidth]{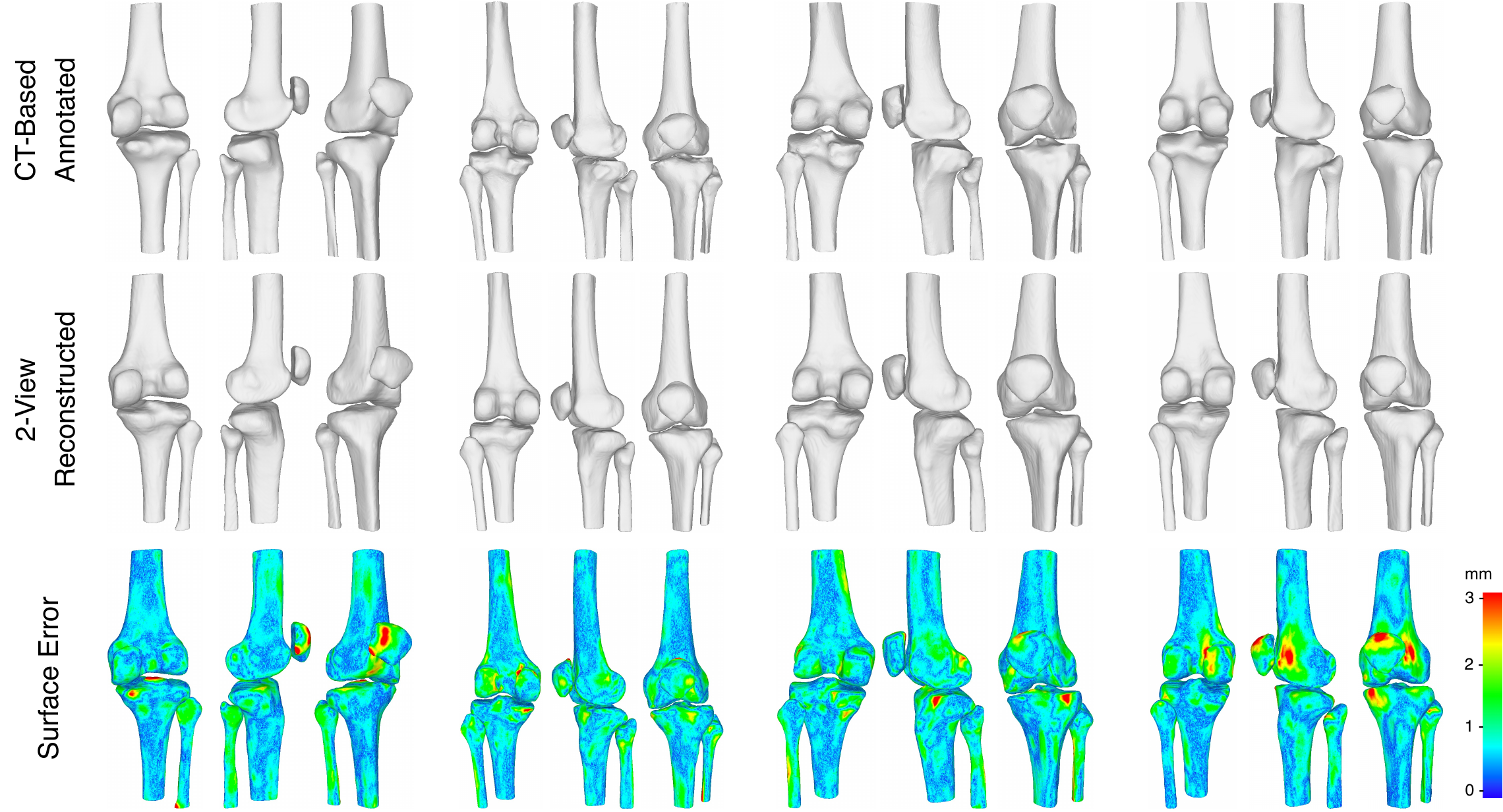}
\caption{
\textbf{Qualitative comparison of two reconstruction methods. }
CT-based annotated bone models are obtained manually from the corresponding CT, while 2-view reconstructed models are automatically generated by our proposed \nickname{} from biplanar X-rays. We also visualize the surface error (mm) between two reconstructed models.
The position, pose, and contour shape of 2-view reconstructed models are exactly consistent with CT-based annotated models. The error is slightly higher in patella and some corner regions because these parts do not appear clearly in the input views due to occlusion.
}
\label{fig:vis_results}
\end{figure}

For qualitative evaluation, we visualized reconstructed bone models in Fig.~\ref{fig:vis_results} to better compare the reconstruction details. Specifically, four cases were visualized, where each case was rendered from three viewpoints, i.e., $0^{\circ}$, $90^{\circ}$, and $180^{\circ}$. The position, pose, and contour shape of 2-view reconstructed models are exactly consistent with CT-based annotated models. For the analysis of reconstructed local details, we also visualized the surface distance (mm) between CT-based annotated and 2-view reconstructed models. The overall distance is lower than 1~mm, while the distance is slightly higher ($\geq$~2~mm) in the patella and some corner regions because these parts do not appear clearly in the input views due to occlusion.

\subsection*{Human-Centric Evaluation: User Study}

\begin{figure}
\centering
\includegraphics[width=1.0\linewidth]{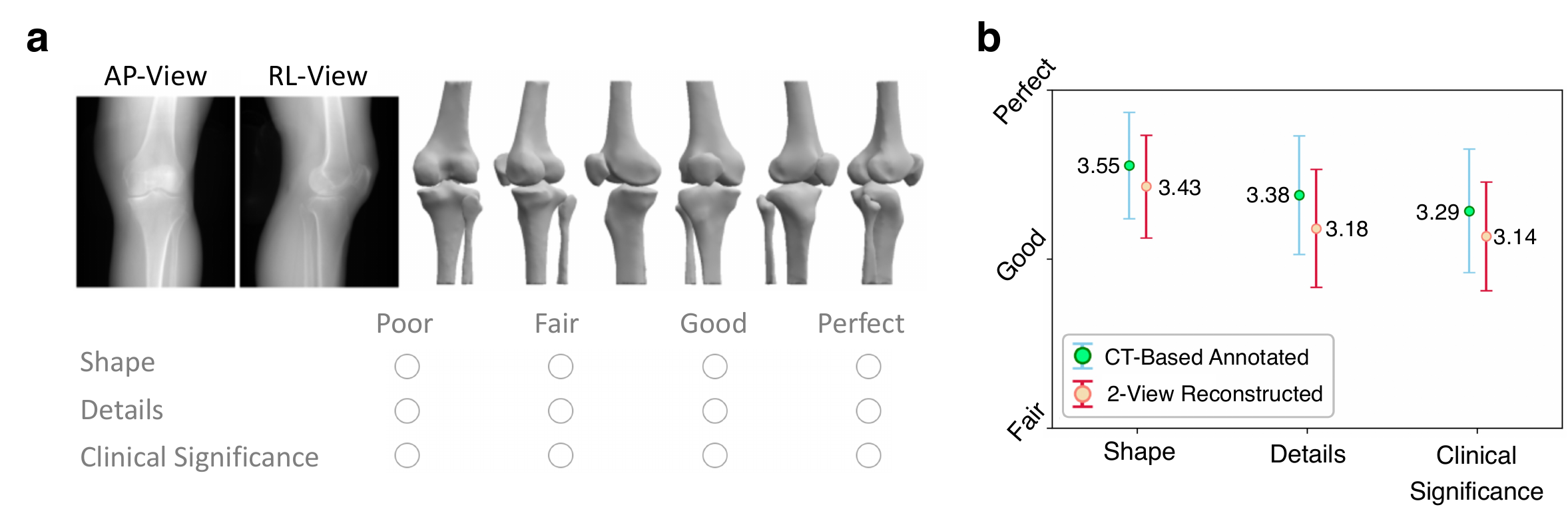}
\caption{
\textbf{User study to compare the clinical difference.}
\textbf{a:} For each case, given 2-view X-ray images as the reference, the user was asked to evaluate bone models (CT-based annotated or 2-view reconstructed) in terms of shape, details, and clinical significance. 
\textbf{b:} Evaluation results (mean and standard deviation) were collected from 10 experts.
}
\label{fig:ques_results}
\end{figure}

We conducted a user study to compare the clinical difference between CT-based annotated bone models and 2-view reconstructed bone models. Specifically, we designed a questionnaire from 40 test cases. For each case, we prepared its 2-view X-rays (i.e., AP and RL views), the corresponding CT-based annotated bone models, and the corresponding 2-view reconstructed bone models. Then, each case involved 2 questions -- 1.) 2-view X-rays and CT-based annotated bone models rendered from six different viewpoints; 2.) 2-view X-rays and 2-view reconstructed bone models rendered from six different viewpoints. An example of questions is shown in Fig.~\ref{fig:ques_results}a, where neither the case ID nor the version of the bone models was disclosed. A total of 80 questions were collected and randomly shuffled. 

For each question, the user was asked to examine the 2-view X-rays and then evaluate the quality of bone models in terms of shape, detail, and clinical significance using four scales (``poor'', ``fair'', ``good'', and ``perfect''). A detailed explanation of four scales is provided in \textit{Methods} section; see \textit{Metrics for Human-Centric Evaluation}. We collected evaluation results from 10 experts, including qualified orthopedists, medical school professors, and medical engineers. Each expert evaluated all cases, assessing a total of 80 bone models (two versions per case). Compared to CT-based annotated models (see Fig.~\ref{fig:ques_results}b), our 2-view reconstructed models obtained comparable scores on all three metrics, which means that experts could hardly identify the difference between CT-based annotated and 2-view reconstructed models given only biplanar X-rays as reference. \cgs{Furthermore, the experts' rating for the clinical significance of the 2-view reconstructed models surpassed a mean score of 3, indicating a consensus on their utility for planning procedures such as HTO.}

\subsection*{Clinical Applicability Evaluation: HTO Simulation}

\begin{figure}
\centering
\includegraphics[width=1.0\linewidth]{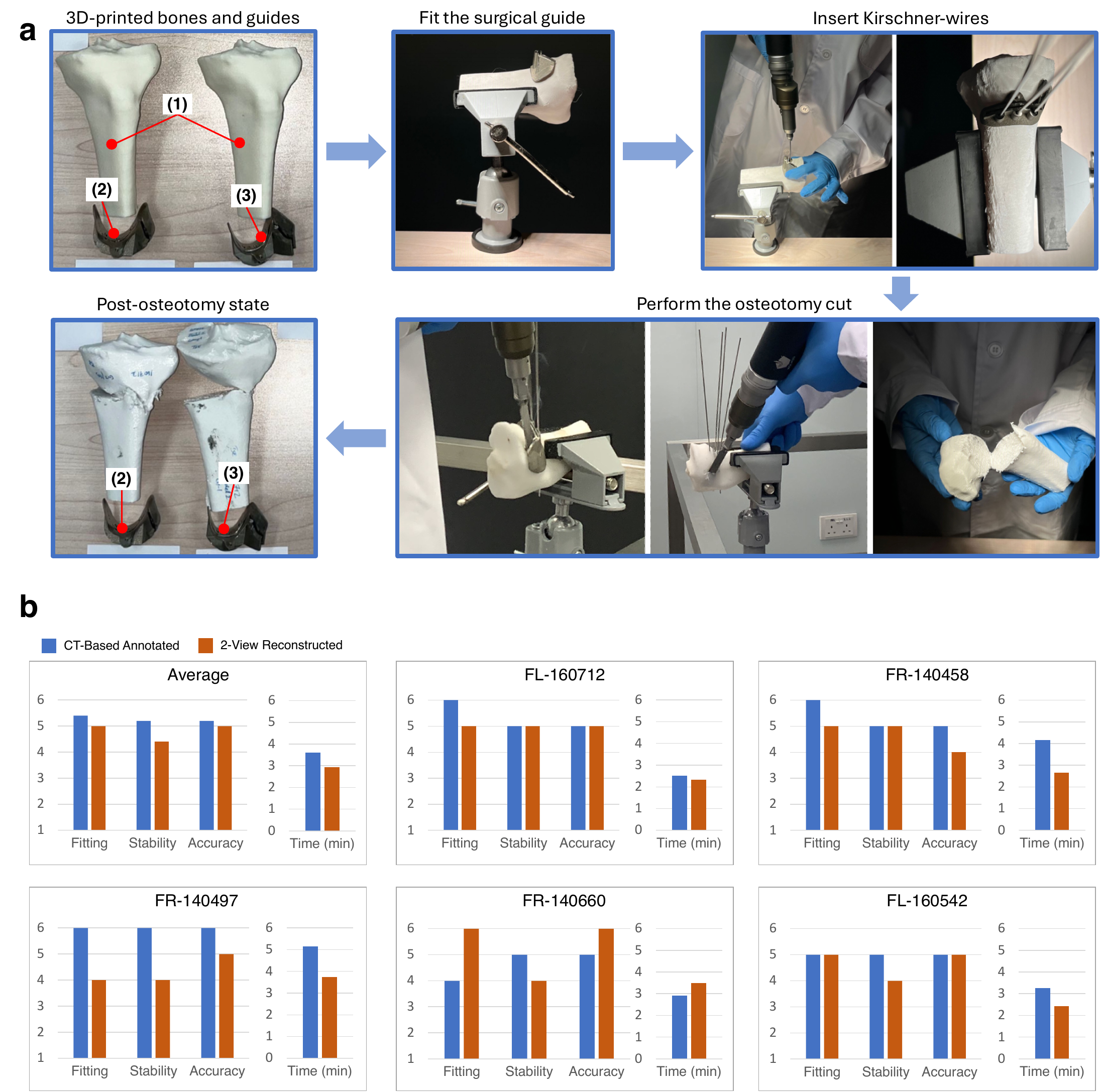}
\caption{
\textbf{HTO simulation to validate the clinical practicality.}
\textbf{a:}
To simulate a HTO, patient-specific bone models were 3D-printed from CT-based annotations to replicate the patient's anatomy (1). Two different versions of surgical guides were then fabricated: one designed from CT-based annotated bone model (2) and another based on 2-view reconstructed bone model (3).
During the simulation, surgical guides were used to direct the procedure. A surgeon would place the guide on the tibial region using a Vise Grip 360° Swivel Head, insert Kirschner-wires through the holes in the surgical guide and into the bone model for fixation purposes, and perform the osteotomy cut using a saw blade.
\textbf{b:}
Three scores (fixing, stability, and accuracy) were used to evaluate the surgical guides, and the operation time (minutes) was also recorded during the operation, demonstrating that 2-view reconstructed bone models have comparable practicality than CT-based annotated models.
}
\label{fig:workshop}
\end{figure}

We invited surgeons to perform a high tibial osteotomy (HTO) simulation to validate the clinical applicability of patient-specific surgical guides designed based on reconstructed bone models. HTO~\cite{wright2005high, mcnamara2013high} is a surgical procedure aimed at realigning the knee joint by cutting and reshaping the tibia to reduce pressure on the damaged cartilage, often used to treat knee osteoarthritis or joint deformities. In HTO, surgical guides are designed to fit the patient's anatomy and ensure precise bone cuts or drill placements~\cite{fayard2024patient, yam20213d}. 
Details of the HTO simulation are illustrated in Fig.~\ref{fig:workshop}a. For each patient, we designed two versions of surgical guides based on CT-based annotated bone models and our 2-view reconstructed bone models using Materialise 3-matic~\cite{materialise3matic}, respectively. Examples of surgical guides are shown in Fig.~\ref{fig:workshop}a, labeled as (2) and (3). In addition to 3D printing two versions of surgical guides, we 3D-printed CT-based annotated bone models to replicate the real patient's anatomy. 

During the HTO simulation, the surgeons evaluated each surgical guide using three scores, fitting, stability, and accuracy, to assess how well the surgical guide conforms to the patient’s anatomy, maintains its position during surgery, and ensures precise guidance for bone cuts, which are essential for achieving successful surgical outcomes~\cite{khan2013surgical, rau2021concept, geiger2023drilling}.
The scores were evaluated on a scale of 1 to 6, based on surgeons' subjective experience. The duration of each HTO operation was also recorded, starting from the placement of the surgical guide on the bone model to the completion of the cut. Detailed explanations on evaluation scores are provided in \textit{Methods} section; see \textit{Metrics for Clinical Application Evaluation}. 
\cgs{HTO simulation was conducted in a blinded manner for the surgeons, which means that during the procedure, the surgeons did not know which version of surgical guide they were using--whether it was designed based on 2-view reconstructed or CT-based annotated bone models.}
Surgical guides of five patients were evaluated, and the results are shown in Fig.~\ref{fig:workshop}b. On average, 1.) surgeons spent less time with surgical guides designed based on 2-view reconstructed bone models compared to those designed based on CT-based annotated bone models; 2.) surgical guides designed based on 2-view reconstructed bone models achieved comparable scores to those designed based on CT-based annotated bone models. This reveals that the bone models reconstructed by \nickname{} from biplanar X-rays have the potential to be used in surgical guide design during HTO preoperative planning, demonstrating the clinical applicability of our proposed \nickname{}.

\subsection*{\cgs{Experimental Analysis}}

\begin{figure}
\centering
\includegraphics[width=1.0\linewidth]{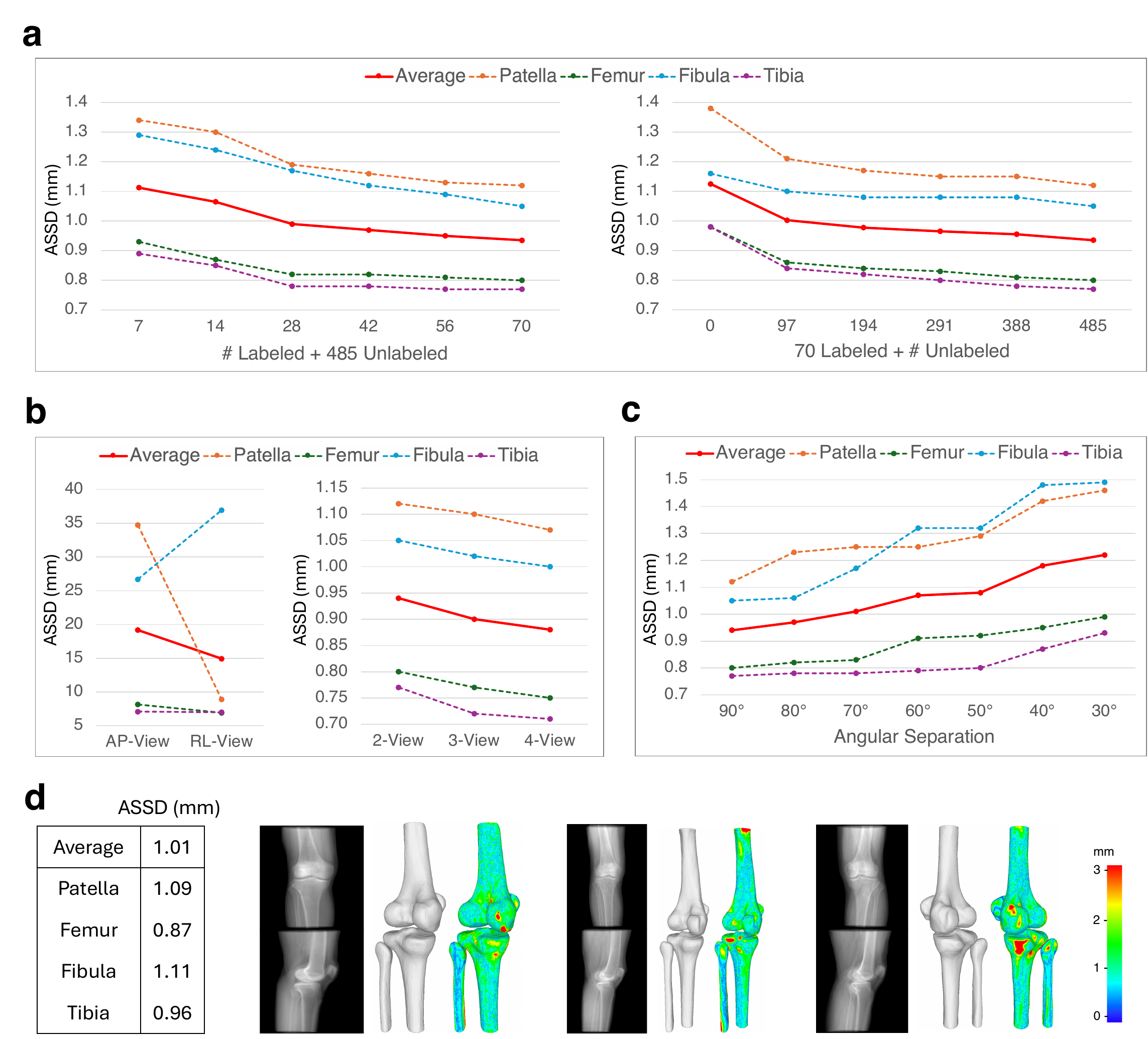}
\caption{
\textbf{Experimental analysis.}
\textbf{a:} Sensitivity analysis of model performance with respect to the number of labeled and unlabeled training pairs.
\cgs{
\textbf{b:} Experiments with different numbers of input views, showing that a 2-view setup provides the optimal trade-off for accurate knee bone reconstruction.
\textbf{c:} Evaluation of model robustness under reduced angular separations, simulating clinical scenarios where gantry rotation is restricted to less than 90°.
\textbf{d:} Assessment of generalization on an external dataset (20 cases), where low ASSD values (mm) and qualitative visualizations confirm the strong performance of SSR-KD on unseen data.
}
}
\label{fig:analysis}
\end{figure}

In this study, we used unlabeled data in the training process as obtaining labeled data is time-consuming and requires expertise to manually annotate bone models from CT. 
To better understand the robustness and sensitivity of our \nickname{} to varying numbers of labeled/unlabeled data, additional experiments were conducted and shown in Fig.~\ref{fig:analysis}a. Specifically, our SSR-KD performs well with as few as 28 labeled data as it also learns knowledge from unlabeled data. However, performance degrades sharply when using either insufficient labeled data (\cgs{$\leq$14}), which fails to model the ground-truth distribution, or limited unlabeled data, which restricts the model's ability to learn diverse anatomical variations.
Our research reveals that \cgs{485 unlabeled data and} only 28 labeled data are sufficient to train a highly competent network capable of reconstructing the femur and tibia with an impressive accuracy of approximately 0.8 mm surface error.

To determine the optimal number of views for bone reconstruction, we conducted additional experiments evaluating SSR-KD with varying numbers of input projections (Fig.~\ref{fig:analysis}b). The results reveal two key findings. First, a single view is insufficient, as severe occlusions prevent the capture of the complete geometry. \cgs{Second, while using more than two views yields slight performance improvements, the gain is significantly less than that achieved when moving from 1-view to 2-view. These experiments demonstrate that a 2-view approach provides the optimal trade-off, offering sufficient complementary spatial information for accurate knee bone reconstruction.}

\cgs{To simulate intraoperative constraints where gantry rotation may be restricted to less than 90°, we evaluated the performance of SSR-KD with reduced angular separations (Fig.~\ref{fig:analysis}c). Although reducing the angular separation inherently decreases the geometric information available for reconstruction, our SSR-KD maintains clinical utility, as the surface error for femur and tibia remains below 1.0 mm even at 30°. However, this information loss has a more significant effect on the patella and fibula. Due to their small size and thin shape, they are more susceptible to being obscured in the projections, leading to a greater increase in surface error as the angle is reduced.}

\cgs{Finally, to validate the generalizability of our model and address potential training set bias, we conducted a cross-validation study using an external dataset. This dataset comprises 20 cases sourced from a different institution and was annotated by a separate group of experts to minimize annotator bias. As shown in Fig.~\ref{fig:analysis}d, our SSR-KD performed effectively on the external out-of-distribution dataset, reconstructing the femur and tibia with an average surface error below 1.0 mm. This result confirms the robust generalizability of our proposed method and training strategy, demonstrating its effectiveness on unseen data from a different clinical environment.}

\section*{Discussion}

This study introduces a semi-supervised reconstruction framework, termed SSR-KD, for generating 3D bone models from biplanar X-ray projections. The primary goal was to develop a method that could produce highly accurate reconstructions. Evaluated on an independent testing set, the SSR-KD framework demonstrated high reconstruction quality, achieving an average surface error below 1 mm.
In comparison to previous bone reconstruction studies (Tab.~\ref{tab:results}), our SSR-KD framework significantly outperforms methods based on SSM~\cite{baka2012statistical} and SSIM~\cite{klima2016intensity}. This superior performance is attributed not only to the powerful learning capabilities of deep learning networks in capturing prior anatomical knowledge, but also to the significant challenges in optimizing SSM/SSIM-based methods.
While some deep learning approaches utilize a volume-based representation~\cite{kasten2020end}, these often suffer from high memory requirements and computational costs, which can lead to low-resolution models and inadequate reconstruction of fine local details.
In contrast, our proposed \nickname{} formulates the reconstruction as the learning of occupancy fields, which is memory-efficient and has proven to achieve the highest reconstruction quality among all compared methods.

\cgss{
In our experiments, while the use of Digitally Reconstructed Radiographs (DRRs) provides a controlled environment for validating the proposed reconstruction algorithm, it relies on a simplified imaging physics model. 
In clinical scenarios, physical nonlinearities such as scatter and beam hardening, along with geometric non-idealities like gantry jitter, are inevitable. These factors introduce complex artifacts (e.g., shading, streaks, and cupping) that can degrade input quality and potentially affect reconstruction accuracy compared to idealized simulations. 
However, modeling all these complex physical interactions in a simulation environment is computationally intensive and non-trivial. Instead, we plan to address this gap in our future work by collecting a large-scale dataset of real clinical X-rays. This will allow us to further develop and fine-tune the model, ensuring its robustness and applicability in real-world clinical settings.
}

In the HTO simulation, the surgical guide designed based on 2-view reconstructed bone models performed better than the CT-based version for the FR-140660 case. We attribute this to minor and unavoidable errors in the 3D printing process~\cite{yeung2020accuracy, msallem2024dimensional, caiti2018positioning}. 
The highly precise CT-based guide, designed for perfect conformity, was paradoxically more sensitive to these small manufacturing deviations, which can lead to a poor physical fit. 
Conversely, the guide designed based on 2-view reconstructed bone models was less detailed, creating natural gaps that tolerate these minor printing inaccuracies. 
We believe the significant error seen in the FR-140660 case was a rare and random event, as the standard and high-precision CT-based approach performed excellently in the other four cases where printing errors were likely less impactful.

\begin{figure}
\centering
\includegraphics[width=1.0\linewidth]{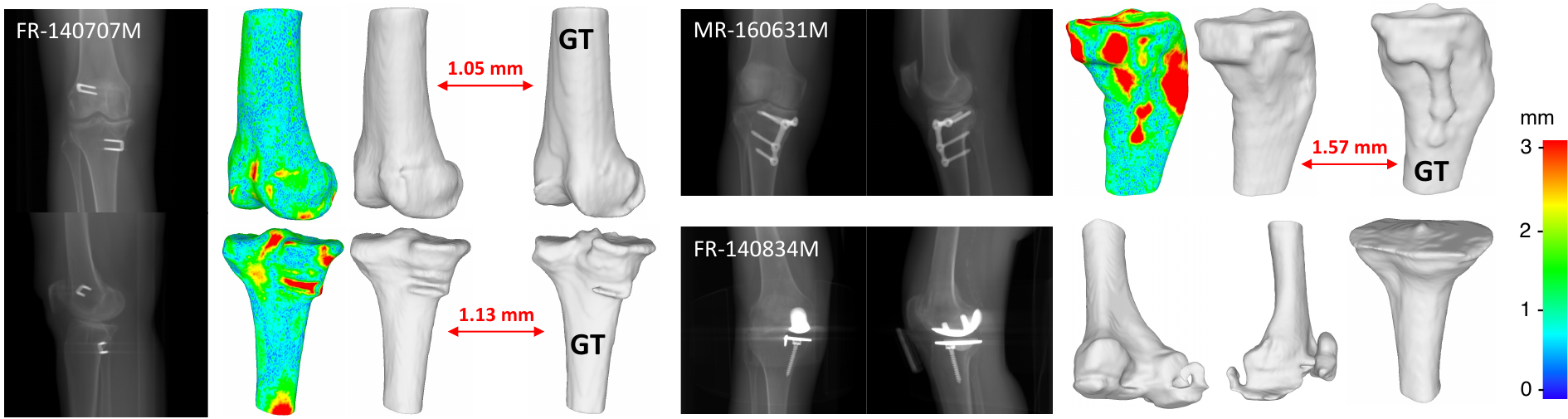}
\caption{
\cgs{
\textbf{Case study on reconstruction for patients with metal implants.}
For FR-140707M (bone staples) and MR-160631M (bone plate and screws) with implants that are small and cause minimal disruption to the bone's structure, the overall shape of the bone was correctly reconstructed. For better comparison, ground-truth (GT) bones are annotated for these two cases, and the surface errors are visualized.
For FR-140834M (partial knee implants) involving large implants that significantly alter the bone's structure, our SSR-KD was unable to reconstruct the bone shape accurately.
}
}
\label{fig:implants}
\end{figure}

To evaluate the performance of our SSR-KD on patients with metal implants, we examined three clinical cases, as shown in Fig.~\ref{fig:implants}. For cases with smaller implants causing minor bone disruption, such as staples (FR-140707M) and screws (MR-160631M), SSR-KD successfully reconstructed the overall bone anatomy, and in some instances, even partially reconstructed the implants. This suggests a strong potential for applications in surgical planning and post-operative assessment. However, for FR-140834M (partial knee implants) with large implants that significantly alter the bone's structure, our SSR-KD failed to produce an accurate reconstruction. To enhance the reconstruction quality for cases with metal implants, it is essential to expand the training dataset with more diverse cases involving metal implants. Furthermore, more advanced algorithms, such as incorporating parameterized templates for the highly standardized shapes of implants, can be explored to enable the separate and accurate reconstruction of both bone and implants. We will pursue this direction in our future work.

\cgs{
The experiments presented in this study focused on the reconstruction of the knee. We believe that our proposed SSR-KD framework can be adapted effectively for the reconstruction of anatomical regions with characteristics similar to the knee, such as the elbow, with only slight modifications. 
However, for other more complex regions like the maxillary bones or the bones of the digits, the suitability of SSR-KD may be limited due to severe occlusion from certain viewpoints or containing numerous small bones. Therefore, achieving clinically useful reconstruction quality for such complex cases will necessitate the development of more advanced architectures, potentially by integrating geometric priors like symmetry or by exploring alternative data representations.
}

In conclusion, this study proposes a novel deep-learning approach \nickname{} to reconstruct high-quality bone models from biplanar X-ray images. Specifically, the network is designed to take X-ray images as input and predict the occupancy value of each point in the 3D space to form the occupancy field. The bone models can be extracted from the occupancy field via Marching Cubes. Compared with human-aid reconstruction from a CT scan, \nickname{} requires the patient to take only X-rays from two orthogonal views, which significantly reduces the radiation dose and enables its use in interventional radiology. We conducted extensive experiments, including quantitative and qualitative analysis, to validate the effectiveness. Furthermore, we held an operating workshop for clinical evaluation, demonstrating that 2-view reconstructed bone models can be the clinical substitutes for those manually annotated from CT. Practically, this study can help to relax the hardware requirement for the reconstruction of bone models, which truly benefits hospitals with limited medical resources.

\section*{Methods} \label{sec:method}

\newcommand{\pred}{5-dim occupancy vector}

\subsection*{Dataset} \label{sec:data}
We collected 605 one-leg CT scans covering the entire knee joint with various ages, genders, and scanning ranges of the knee. The dataset specifically comprises 186 female-left, 146 female-right, 142 male-left, and 131 male-right legs, with the field of view varying between 145$\times$145$\times$108~mm to 375$\times$375$\times$509~mm, showing the considerable diversity in the data distribution. Among them, 120 CT scans were manually annotated with bone models (patella, femur, fibula, and tibia) by three orthopedic experts, and the remaining 485 CT scans are considered as unlabeled data.
CT scans were sampled to the resolution of $256^3$ with the voxel-wise spacing ranging from (0.42, 0.57, 0.57)~mm to (1.98, 1.46, 1.46)~mm. Annotated bone models were normalized to have coordinates ranging from -1 to +1, and then simplified to have around 8,000 points.
We followed previous work~\cite{shen2019patient} to generate anterior-posterior (AP) and right-left (RL) projections from CT scans by digitally reconstructed radiograph (DRR). This process was implemented using TIGRE package with a cone-beam X-ray propagation model. \cgss{To simulate non-ideal conditions, we modeled statistical quantum noise assuming an incident photon flux of $1 \times 10^5$ photons per pixel, and incorporated electronic noise using a Gaussian distribution ($\mu=0, \sigma=10$).}
Finally, we have 120 labeled (DRR, CT, bone models) pairs and 485 unlabeled (DRR, CT) pairs. Labeled pairs were split into 70 for training, 10 for validation, and 40 for testing. Therefore, in our experiments, 70 ($\sim$12.6\%) labeled data and 485 ($\sim$87.4\%) unlabeled data were utilized for model training.

\subsection*{Problem Formulation}

Given 2-view X-ray projections or a CT scan, the goal is to reconstruct the underlying 3D geometry of the target bones. For a single bone model, we assume that it is watertight and represent its 3D surfaces as an implicit function $f(\cdot)$. Specifically, for a point $p \in \mathbb{R}^3$ in the 3D space, $f(\cdot)$ estimates the occupancy value $s$ of $p$:
\begin{equation}
s = f(I, p)=\left\{\begin{array}{ll}
1, & \text{if $p$ is inside the surface,} \\
0, & \text{otherwise,}
\end{array}\right.
\end{equation}
where $I$ is the input (X-rays or CT). In this study, we aim to simultaneously reconstruct four bone models, including patella, femur, fibula, and tibia. Observing that these bone models do not overlap, we extend the above formulation and reformulate the implicit function as
\begin{equation}
    \begin{split}
        & x = \pi(p),\ F = g(I), \\
        & F|_x = \text{Interp}(F, x), \\
        & s = f(I, p) = f(F|_x, p) \in \mathbb{R}^5,
    \end{split}
\end{equation}
where $\pi(\cdot)$ is the projection function that transforms $p$ from the world coordinate system to the pixel/voxel coordinate system; $g(\cdot)$ is the encoder network to extract semantic features $F$ from the input $I$; $\text{Interp}(\cdot)$ represents interpolating pixel/voxel-aligned features $F|_x$ from $F$ at position $x$.
Differently, we represent the occupancy $s$ as a 5-dimensional vector, where the 5 values (0-4) indicate the probability of $p$ inside patella (1), inside femur (2), inside fibula (3), inside tibia (4), and outside (0), respectively. We denote $s_i$ as $i^\text{th}$ element of $s$. The ground-truth occupancy $s$ is one-hot since there is no overlapping. In the inference stage, the predicted occupancy vector $\hat{s}$ is a probability vector, satisfying $\sum_i \hat{s}_i = 1$. To obtain the one-hot occupancy vector, the highest value of $\hat{s}$ will be set to 1, and others are set to 0; i.e., $\hat{s} = \text{one-hot}(\hat{s})$.

The reconstruction of bone models is shown in Fig.~\ref{fig:inference}a. A dense set (i.e., $256^3$) of points are uniformly sampled from $[-1, +1]^3$, and then the network will predict their 5-dimensional occupancy vectors. Hence, the occupancy field of $i^\text{th}$ ($i \geq 1$) bone model can be formed by all $\hat{s}_i$, and then the iso-surfaces are extracted using Marching Cubes algorithm~\cite{lorensen1987marching}.

\subsection*{Reconstruction Networks} \label{sec:recon}

\begin{figure}
\centering
\includegraphics[width=1.0\linewidth]{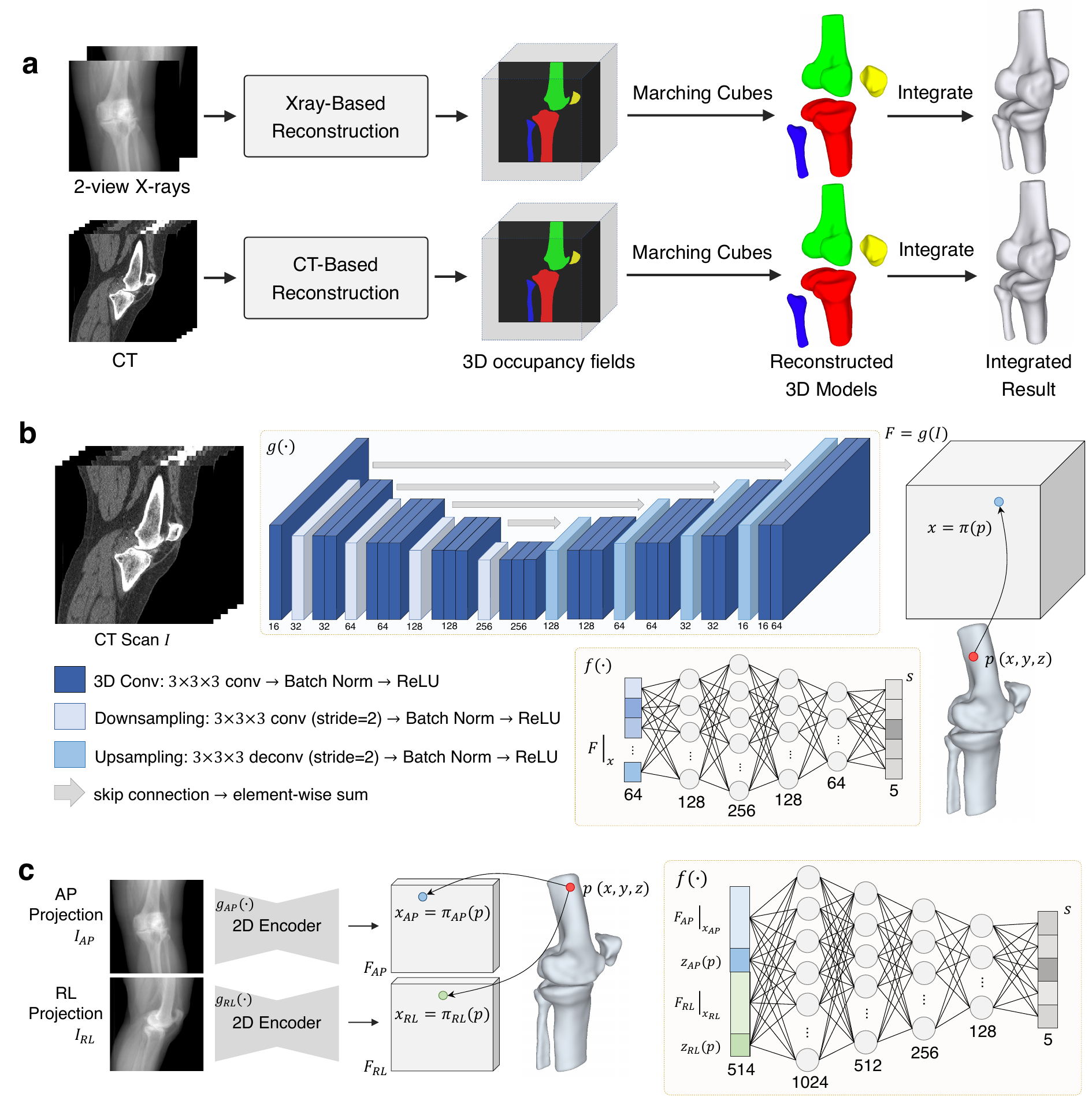}
\caption{
\textbf{Reconstruction pipelines and network designs for different inputs.}
\textbf{a:} Bone model reconstruction. The occupancy field of each bone can be estimated from a CT scan or 2-view X-rays. The 3D model is extracted from the occupancy field using Marching Cubes~\cite{lorensen1987marching}. Multiple bones are integrated as the final reconstructed result. 
\textbf{b:} Reconstruction from a CT scan. For the input CT scan, a V-Net~\cite{milletari2016v} is used as the encoder for 3D feature extraction. The feature of an arbitrary point can be interpolated from the 3D feature map to estimate its occupancy value.
\textbf{c:} Reconstruction from 2-view X-ray images. For two input X-ray images, two stacked hourglass networks~\cite{newell2016stacked} are used as encoders for feature extraction. The feature of an arbitrary point can be interpolated from two 2D feature maps to estimate its occupancy value.
}
\label{fig:inference}
\end{figure}

In Fig.~\ref{fig:inference}b, we represent a network for the reconstruction from a CT scan (3D input). For a 3D CT scan $I \in \mathbb{R}^{W\times H\times D}$, we employ V-Net~\cite{milletari2016v} as the 3D feature encoder $g(\cdot)$. For a 3D point, the projection is defined as $x = \pi(p) = p$ and the implicit function $f(\cdot)$ takes the voxel-aligned feature $F|_x \in \mathbb{R}^{64}$ as the input for predicting the 5-dimensional occupancy vector $s = f(F|_x)$. In practice, $f(\cdot)$ is implemented with multi-layer perceptrons, and the channels are 64$\rightarrow$128$\rightarrow$256$\rightarrow$128$\rightarrow$64$\rightarrow$5.

In Fig.~\ref{fig:inference}c, we design a reconstruction for X-rays (2D) as the input. Specifically, the inputs are 2-view (AP and RL) X-ray projections $I_\text{AP} \in \mathbb{R}^{W\times H}$ and $I_\text{RL} \in \mathbb{R}^{W\times H}$. For each input view $v \in \{\text{AP}, \text{RL}\}$, we employ an independent (not shared) stacked hourglass network~\cite{newell2016stacked} as the 2D feature encoder $g_v(\cdot)$. The projection function $\pi_v: \mathbb{R}^3 \rightarrow \mathbb{R}^2$ is defined as the same as DRR to project a 3D point to the 2D imaging plane of $v$-view. For a 3D point $p$, we calculate its projected coordinates $x_v = \pi_v(p)$ and then obtain the pixel-aligned feature $F_v|_{x_v}$ from $v$-view. In addition, we utilize the distance $z_v(p)$ of $p$ to the imaging plane of $v$-view to provide depth information. Finally, pixel-aligned features and depth values are concatenated as
$[F_\text{AP}|_{x_\text{AP}}; z_\text{AP}(p); F_\text{RL}|_{x_\text{RL}}; z_\text{RL}(p)] \in \mathbb{R}^{(256+1)\times 2}$.
The implicit function $f(\cdot)$ takes the concatenated features as the input and predicts the 5-dimensional occupancy vector $s$. In practice, $f(\cdot)$ is implemented with multi-layer perceptrons, where the channels are  514$\rightarrow$1024$\rightarrow$512$\rightarrow$256$\rightarrow$128$\rightarrow$5.

\subsection*{Semi-Supervised Reconstruction Framework with Knowledge Distillation (\nickname{})} \label{sec:semi}

In this study, both labeled and unlabeled data pairs are used for semi-supervised training. Specifically, the proposed \nickname{} training framework is composed of two steps, including the training of the CT-based reconstruction network $\mathcal{N}_\text{CT}$ with labeled data, and the training of the Xray-based reconstruction network $\mathcal{N}_\text{Xray}$ with both labeled and unlabeled data.

In the first step, we train the CT-based reconstruction network ($\mathcal{N}_\text{CT}$) with labeled data, because CT scans can provide detailed spatial information, 80 labeled data pairs [CT, bone models] are sufficient for training a CT-based network to achieve good performance. During training, 5,000 points are uniformly sampled from the 3D space, and the cross-entropy loss is computed based on the predicted and ground-truth occupancy vectors. The network is trained for 400 epochs with a batch size of 3 and optimized by a momentum gradient descent optimizer with a momentum of 0.98 and an initial learning rate of 0.01 decayed by 0.1 per 100 epochs.

\begin{figure}[t]
\centering
\includegraphics[width=1.0\linewidth]{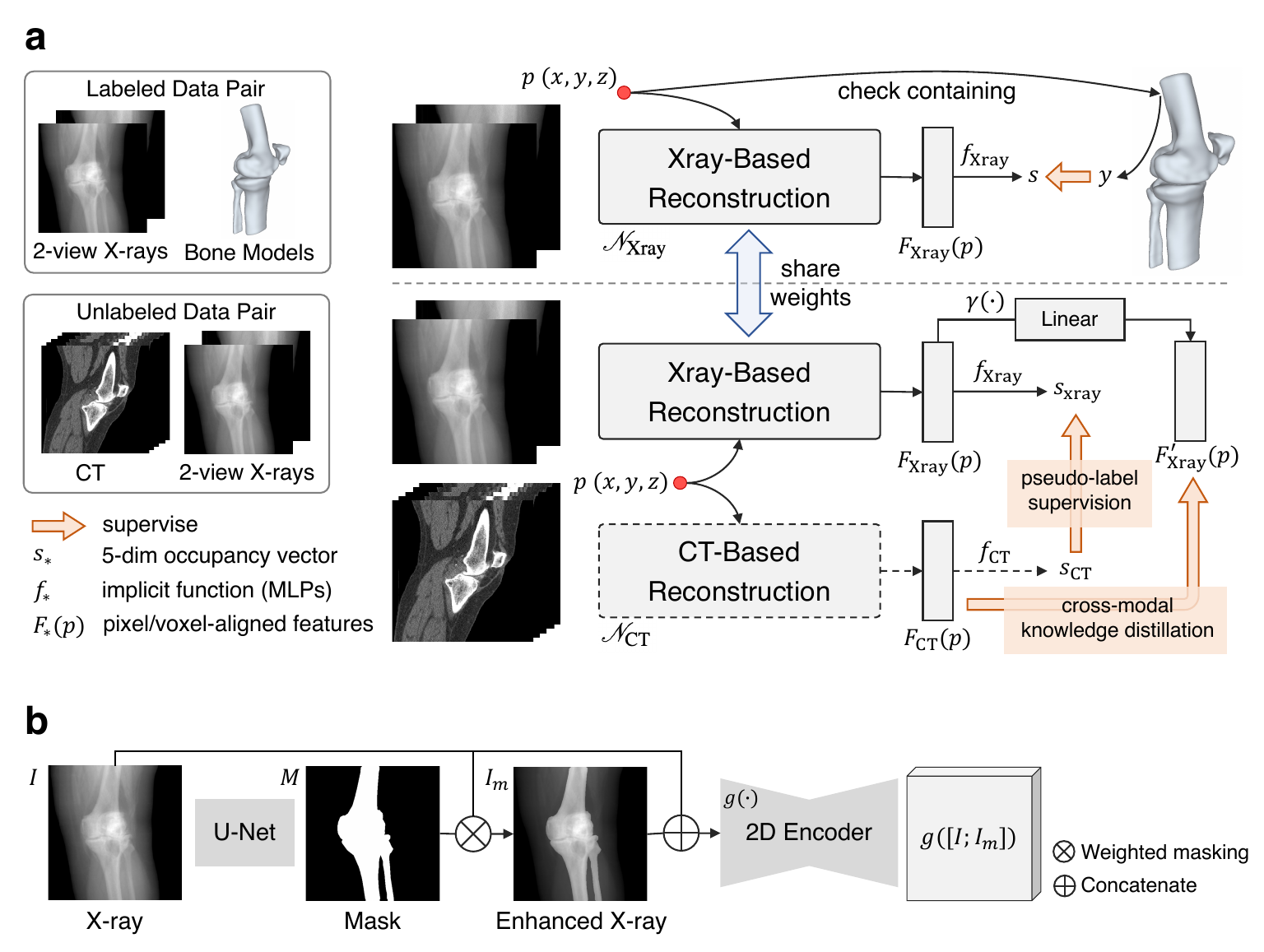}
\caption{
\textbf{SSR-KD and input enhancement.}
\textbf{a:} The SSR-KD training framework for X-ray-based reconstruction network. Both labeled and unlabeled data pairs are used during training. For labeled data, the output is supervised by the ground-truth occupancy value; for unlabeled data, a well-trained CT-based reconstruction network is used for pseudo-label supervision and cross-modal knowledge distillation. 
\textbf{b:} Input with enhanced X-rays. For each input view, a 2D segmentation network (U-Net) is applied to generate the foreground (bones) mask for synthesizing the enhanced X-ray image by weighted masking. The enhanced X-ray is then concatenated with the original X-ray as the new input for the 2D encoder $g(\cdot)$ of X-ray-based reconstruction network.}
\label{fig:framework_main}
\end{figure}

In the second step, we train the X-ray-based network ($\mathcal{N}_\text{Xray}$), which is more challenging due to partial occlusion and insufficient labeled data. To address this, we propose a novel framework \nickname{} by leveraging semi-supervised learning and knowledge distillation. The overview of \nickname{} is shown in Fig.~\ref{fig:framework_main}a.
Specifically, for a labeled data pair and a point $p \in \mathbb{R}^3$, the ground-truth occupancy vector $y$ is obtained by checking the containment of $p$ in annotated bone models. Hence, the predicted $s$ is supervised by $y$ using cross-entropy (CE) loss:
\begin{equation}
    \loss_\text{labeled} = \loss_\text{CE}(s, y).
\end{equation}
For an unlabeled data pair and a point $p$, the Xray-based network $\mathcal{N}_\text{Xray}$ takes 2-view X-ray projections as the input and produces $p$'s combined pixel-aligned features $F_\text{Xray}(p) \in \mathbb{R}^{514}$ and the occupancy vector $s_\text{Xray}$. To perform semi-supervised learning, we adopt the CT-based network $\mathcal{N}_\text{CT}$ trained in step-(1) to generate $p$'s voxel-aligned features $F_\text{CT}(p) \in \mathbb{R}^{64}$ and pseudo occupancy vector $s_\text{CT}$ from the input CT scan. The pseudo label is defined as $y_\text{pseudo} = \text{one-hot}(s_\text{CT})$ and the pseudo-label supervision is given by
\begin{equation}
    \loss_\text{pseudo} = \loss_\text{CE}(s_\text{Xray}, y_\text{CT}).
\end{equation}
Furthermore, to enhance the learning of pixel-aligned feature representation, we introduce cross-modal knowledge distillation (KD) as follows. Firstly, a learnable linear layer $\gamma(\cdot)$ (514$\rightarrow$64) is applied to $F_\text{Xray}(p)$ for dimension alignment, then the proposed knowledge distillation is formulated as
\begin{equation}
\begin{split}
    & F_\text{Xray}^{'}(p) = \gamma(F_\text{Xray}(p)), \\
    & \loss_\text{kd} = \big\|F_\text{Xray}^{'}(p) - F_\text{CT}(p)\big\|_1,
\end{split}
\end{equation}
where $\|\cdot\|_1$ represents $l_1$-norm. Hence, the loss for unlabeled data pairs is defined as
\begin{equation}
    \loss_\text{unlabeled} = w_u \cdot \loss_\text{pseudo} + w_k \cdot \loss_\text{kd},
\end{equation}
where $w_u \in [0, 1]$ and $w_k \in \mathbb{R}$ are balancing weights. During \nickname{} training, we sample a batch of data composed of half labeled and half unlabeled data pairs as the input for mini-batch training. The joint loss function is defined as
\begin{equation}
\begin{split}
    \loss &= (1 - w_u) \cdot \loss_\text{labeled} + \loss_\text{unlabeled} \\
          &= (1 - w_u) \cdot \loss_\text{labeled} + w_u \cdot \loss_\text{semi} + w_k \cdot \loss_\text{kd}.
\end{split}
\label{eq:joint}
\end{equation}
In practice, the network is trained for 400 epochs with a batch size of 4 and optimized by a momentum gradient optimizer with a momentum of 0.98 and an initial learning rate of 0.01 decay by 0.1 per 100 epochs. $w_u$, $w_k$ in Eqn.~\ref{eq:joint} are empirically set to 0.5 and 0.1, respectively. During training, for labeled data, \cgss{2,500 points are uniformly sampled from the 3D space, and another 2,500 points are sampled near the bone surface. Specifically, these near-surface points are generated by applying Gaussian noise ($\mu=0, \sigma=1$ mm) to points on the bone surface.} For unlabeled data, 5,000 points are uniformly sampled from the 3D space. 
\cgss{
During inference, we reconstruct at a resolution of $256^3$. To mitigate memory constraints, the $\sim$16.8 million query points are processed sequentially in mini-batches of 50,000.
}

\subsection*{Input With Enhanced X-ray Projection}

To further improve the reconstruction quality, we propose to highlight the reconstruction targets by enhancing the intensity of foreground areas (bones) and suppressing the intensity of background areas (soft tissues). Specifically, as shown in Fig.~\ref{fig:framework_main}b, taking an X-ray image $I$ as the input, a segmentation network (U-Net~\cite{ronneberger2015u}) is used to generate a binary mask $M$ (0: background, 1: foreground). \cgss{The U-Net design is highly robust to predict high-quality masks, achieving 96.9\% segmentation mIoU on our test set.} Then, the enhanced image $I_m$ is calculated by
\begin{equation}
    I_m = I \odot M + w_m \cdot I \odot (E - M),
    \label{eq:enhance}
\end{equation}
where $w_m \in [0, 1]$ is the scaling weight, $\odot$ is element-wise multiplication, and $E$ is an all-one matrix of the same size as $I$ and $M$. Finally, we concatenate $I$ and $I_m$ as the new input $[I; I_m]$ for the 2D encoder of X-ray-based reconstruction network. \cgss{This design ensures that raw information is always preserved, preventing data loss from potential segmentation errors, while $I_m$ serves as an additional cue to guide the network's attention. Empirically, this design proves effective, reducing the overall reconstruction error (ASSD) by 0.12 mm.} The segmentation network is trained on 300 annotated data pairs [X-ray, mask] with a batch size of 8 for 400 epochs and optimized by a momentum of 0.98 and an initial learning rate of 0.01 decay by 0.1 per 100 epochs. $w_m$ in Eqn.~\ref{eq:enhance} is empirically set to 0.5. Note that the segmentation network is pre-trained and then fixed during the training of X-ray-based reconstruction network.

\subsection*{Metrics for Quantitative Evaluation} \label{sec:eval_quan}

Given a reconstructed bone model $\hat{B}$ and the reference bone model (ground-truth) $B$, we denote $\hat{P}$ and $P$ as the points in the surface of $\hat{B}$ and $B$, respectively. Therefore, average symmetric surface distance (ASSD) and Hausdorff distance (HD) are defined as
\begin{equation}
\begin{split}
    & d(x, Y) = \min_{y \in Y} \|x - y\|_2, \\
    & \text{ASSD}(\hat{P}, P) = \frac{1}{2} \Big(
    \text{Mean}\big\{d(\hat{p}, P)\ \big|\ \forall \hat{p} \in \hat{P}\big\} 
    + 
    \text{Mean}\big\{d(p, \hat{P})\ \big|\ \forall p \in P\big\}
    \Big), \\
    & \text{HD}(\hat{P}, P) = \max \Big\{
    \max_{\hat{p} \in \hat{P}}\ d(\hat{p}, P),\ 
    \max_{p \in P}\ d(p, \hat{P})
    \Big\}.
\end{split}
\end{equation}
In practice, we uniformly sample 16,348 points from each bone model as finite point sets to approximate $\hat{P}$ and $P$ for the above calculation. In addition, we voxelize the bone models into binary volumes (0: bone; 1: others) $\hat{V}$ and $V$ for the Dice similarity coefficient (DSC, \%) calculation using the definition of true positive (TP), false positive (FP), and false negative (FN), given by
\begin{equation}
    \text{DSC}(\hat{V}, V) = \frac{2\text{TP}}{2\text{TP} + \text{FP} + \text{FN}}.
\end{equation}

\subsection*{Metrics for Human-Centric Evaluation} \label{sec:eval_user}
There are four scales (1 - 4: ``poor'', ``fair'', ``good'', and ``perfect'') to evaluate the reconstructed bone model in terms of shape, details, and clinical significance. Specifically,
\textbf{Shape} and \textbf{Details} are used to measure the correctness of reconstructed global shape and local details, respectively. A higher shape/details score indicates that the shape/details of the model better match the given images. Four scores represent (1) more than 30\%, (2) 10\% - 30\%, (3) 5\% - 10\%, and (4) less than 5\% of shape/details are not correctly reconstructed.
\textbf{Clinical Significance} is used to assess the significance of reconstructed models in clinical scenarios. A higher score indicates that the reconstructed model has broader clinical applications. Four scores represent the reconstructed model (1) cannot be used for any clinical practice, or can be used for (2) only diagnosis, (3) the planning of osteotomies in certain bones, like high tibial osteotomy (HTO), (4) the planning of most osteotomies, like total knee replacement, etc.

\subsection*{Metrics for Clinical Applicability Evaluation} \label{sec:eval_clinical}
\cgs{To assess the clinical applicability of the reconstructed bone models, a High Tibial Osteotomy (HTO) simulation was performed for five patient cases. For each case, a patient-specific surgical guide was designed based on the reconstructed bone models. These guides were used to fit securely onto the bone surface, with slots that precisely direct the osteotomy saw blade according to the pre-operative surgical plan.}
When designing surgical guides, particularly for orthopedic procedures, three critical criteria must be evaluated: \textbf{Fitting}, \textbf{Stability}, and \textbf{Accuracy}. Each of these metrics plays a vital role in ensuring successful surgical outcomes. A well-designed surgical guide should achieve high scores in all three areas, as they are crucial indicators of its effectiveness in guiding the surgeon to perform precise cuts and ensuring the overall success of the procedure.

Surgical guides are often designed using preoperative imaging techniques, such as CT or MRI scans, ensuring a precise match to the patient’s anatomy. For example, custom guides have demonstrated significant reductions in maximum deviation from preoperative plans, achieving deviations of only 2 mm compared to 9 mm for manual techniques~\cite{khan2013surgical}. The \textbf{Fitting} score of a surgical guide refers to how well the device conforms to the anatomical structures of the patient. A well-fitted guide ensures secure placement on the bone surface, which is essential for accurate guidance during surgery. Specifically, the fitting score measures the percentage of surface contact between the surgical guide and the bone model. Six scales are used to assess this: (1) less than 10\% contact, (2) 10\% - 30\% contact, (3) 30\% - 50\% contact, (4) 50\% - 70\% contact, (5) 70\% - 90\% contact, (6) 90\% - 100\% contact.

A stable guide must maintain rigidity under surgical forces. For instance, the stiffness of components like base plates has been shown to significantly affect accuracy during drilling operations~\cite{rau2021concept}. Features such as alignment pins and holes for securing the guide are critical for maintaining stability throughout the procedure. These are particularly important for moldable surgical guides that require precise placement and stabilization~\cite{rau2021concept}. We use \textbf{Stability} score to measure how well the surgical guide stays in place during the cutting or drilling process. Stability is critical, as any movement or shifting can result in inaccuracies and compromise the procedure's success. Specifically, the stability score is typically assessed by applying force or pressure to the guide and measuring the loss of contact between the guide and the bone. The six scales for stability evaluation are: (1) more than 90\% loss of contact, (2) 60\% - 90\% loss of contact, (3) 30\% - 60\% loss of contact, (4) 10\% to 30\% loss of contact, (5) 1\% - 10\% loss of contact, (6) less than 1\% loss of contact.

Studies have demonstrated that using custom surgical guides can significantly improve accuracy. For example, in a cadaveric study, custom-guide-assisted techniques resulted in no violations of accepted error thresholds when aiming for cuts within 4 mm of ideal resection lines~\cite{khan2013surgical}. It is important to identify potential sources of error, such as manufacturing tolerances and user errors during assembly or adjustment. For example, manual adjustments in guide length settings can introduce inaccuracies if not carefully calibrated~\cite{rau2021concept, geiger2023drilling}. The \textbf{Accuracy} score is perhaps the most critical criterion, as it directly impacts surgical outcomes by ensuring that cuts or resections are performed as planned. Specifically, the accuracy score measures how well the surgical guide helps the surgeon achieve precise cuts in terms of position and angle. This is evaluated based on the percentage of good vertical and rotational fitting achieved, as well as the quality of guidance provided by the guide: (1) does not provide any, or provided a (2) minimal, (3) satisfactory, (4) acceptable, (5) accurate, (6) precise guide for the surgeon to make precise cuts and has (1) less than 10\%, (2) 10\% - 30\%, (3) 30\% - 50\%, (4) 50\% - 70\%, (5) 70\% - 90\%, (6) more than 90\% good vertical and rotational fitting.

\section*{\cgs{Data Availability}}
\cgs{The datasets generated and/or analyzed during the current study are not publicly available because this would compromise the patient confidentiality and privacy agreement with the data-providing hospitals, which prohibits any form of public distribution. The minimal dataset that would be necessary to interpret, replicate, and build upon the findings reported in the article is available from the corresponding authors upon reasonable request.}

\section*{Code Availability}
The source code for this study is publicly available on GitHub at \href{https://github.com/xmed-lab/SSR-KD}{https://github.com/xmed-lab/SSR-KD}, including the source code for implementing SSR-KD framework, data generation, and experimental analysis. The code is released under the MIT license. We implemented the network design, model training, and model evaluation using PyTorch~\cite{paszke2019pytorch}. Key dependencies include TIGRE for X-ray projection simulation and PyTorch3D for calculating chamfer distance. Other significant libraries used are Open3D, NumPy, Trimesh, skimage, and SimpleITK. \cgss{The released codebase was tested on a workstation equipped with two NVIDIA GeForce RTX 3090 GPUs (24 GB), Intel Xeon Gold 5218 CPU @ 2.30 GHz, and 128 GB of RAM.}

\section*{Acknowledgements}
This work was supported by a research grant from the Hong Kong Innovation and Technology Fund (Project PRP/041/22FX) \cgs{and partially supported by the Natural Science Foundation of Zhejiang Province (No. LZ23A050002) and the National Natural Science Foundation of China (No. 12175012).}

\section*{Author Contributions}
X. Li, W. Zhao, Y. Lin, and H. Sun conceptualized and designed the study. Y. Lin implemented the method and contributed to the manuscript writing, reivision, and the analysis of the results. H. Sun, W. Yau, and M. Amangeldy designed and organized the HTO simulation to evaluate its clinical applicability. \cgs{Y. Li, R. Aslam, T. Cheng, and V. Le Meur coordinated the data annotation. L. Tse, C. Chui, and K. Cho assessed the annotation quality and provided guidance. Y. Ye and J. Zou developed the manuscript outline and guided its preparation.} All authors read and approved the manuscript.

\section*{Competing Interests}
The authors declare no competing interests.

\vspace{12pt}
\noindent
Received: January 12, 2024; Accepted: XX XXXXXX XXXX; Published online: XX XXXXXX XXXX.

\thispagestyle{empty}
\bibliography{bib}

\newpage

\begin{table}[t]
\centering
\setlength{\tabcolsep}{13.5pt}
\caption{
\textbf{Quantitative evaluation on reconstruction quality. }
Four bones (patella, femur, fibula, and tibia) of a single leg are evaluated. Reconstructed results are compared with bone models manually annotated from the corresponding CT for metric calculation. For distance metrics (ASSD and HD), lower values indicate better quality; for DSC, higher values indicate better quality. \cgss{We also list ASSD results reported in previous studies; note that these values are cited directly from the original literature and were not re-implemented on our dataset.}} \label{tab:results}
\vspace{-2mm}
\begin{tabular}{l|l|c|cccc}
\toprule[1.2pt]
Method & Metric & Average & Patella & Femur & Fibula & Tibia \\ \hline \hline
\multirow{3}{*}{\nickname{} (\textit{ours})} & DSC (\%) $\uparrow$ & 90.9 & 87.5 & 96.1 & 84.1 & 96.0 \\ \cline{2-7}
 & HD (mm) $\downarrow$ & 2.76 & 3.30 & 2.20 & 2.87 & 2.06 \\ \cline{2-7}
 & ASSD (mm) $\downarrow$ & 0.94 & 1.12 & 0.80 & 1.05 & 0.77 \\ \hline
Baka et al. 2012~\cite{baka2012statistical} (SSM) & ASSD (mm) $\downarrow$ & - & - & 1.48 & - & - \\ \hline
Klima et al. 2016~\cite{klima2016intensity} (SSIM) & ASSD (mm) $\downarrow$ & - & - & 1.18 & - & - \\ \hline
Kasten et al. 2020~\cite{kasten2020end} (CNNs) & ASSD (mm) $\downarrow$ & 1.30 & 1.71 & 1.08 & 1.22 & 1.18 \\
\bottomrule[1.2pt]
\end{tabular}
\end{table}


\clearpage 

\end{document}